\documentclass[granma]{svjour} 
\usepackage{graphics,epsfig,citesort} 

\begin{document}  

\bibliographystyle{plain} 

\title{Shakedown of unbound granular material}

\author{ R. Garc\'{\i}a-Rojo and H. J. Herrmann}
\institute{
	Institute for Computer Applications 1, \\ 
           Pfaffenwaldring 27, \\ 
           70569 Stuttgart, GERMANY\\  
           e-mail: rgrojo@ica1.uni-stuttgart.de\\
}
\date{\today}
\maketitle

\begin{abstract}
Compacted unbound granular materials are extensively used as sub-layer in pavement design. Most pavement design guides assume that they are responsible for the degradation and deformation of the roads and railways that they support. Biaxial tests are usually employed to investigate the elasto-plastic response of these materials to cyclic loading. A particularly interesting question is whether a limit load exists, below which the excitations shake down, in the sense that the material does not accumulate further deformations. We have carried out a detailed study of the elasto-plastic behavior of a simple model of unbound granular matter submitted to cyclic loading. The dissipated energy through out the simulation has been used for the characterization of the different regimes of responses. 
\end{abstract}

\section{Introduction.}
Traditional pavement design methods are still almost completely empirical. Long term experience with the performance of on-service roads is supplemented with the results obtained from especially constructed test pavements. Changes in material properties, the loading or the environmental conditions can be hardly investigated. The information obtained from the experiments is therefore limited. The disadvantages of traditional design have become more obvious during the last decades as a consequence of the growth of transportation needs and the urge for using recycled materials. Mechanistic and analytical design procedures have consequently been developed based on the analysis of the response of the structure under specific loads. Understanding the behavior of the components of the structure is of course a pre-requisite of this approach. In this respect, the deformation behavior of unbound granular materials (basic component of roads and pavements) has been one of the main topics of pavement engineering for many years \cite{taciroglu02,lekarp00,sharp84}, since they are principal responsible of the rutting and cracking of the pavement \cite{austroad92}. 

Grain scale investigation of cyclic loading of granular materials is possible using discrete element methods, that solve the dynamical evolution of the system according to the interaction between the particles. The quasi-static Coulomb's friction on lasting contacts has been extensively studied by means of algorithms of Molecular  Dynamics (MD) \cite{cundall79}. In this method, a visco-elastic interaction is introduced in each contact. The evolution of the grains is solved explicitly by the numerical solution of the equations of motion. The grains are usually represented  by spheres (in 3D) or disks (in 2D). One expects that this approximation will reproduce the behavior of smooth-grained unbound granular material (UGM), such as gravels. Materials like crushed aggregates, composed by grains with more irregular shapes, have also been modeled \cite{cundall89,alonso03,tillemans95,kun96b}, leading to more complicated and therefore less efficient algorithms.

It is usually assumed that the global response of an UGM to cyclic loading can be decomposed into resilient deformation, and permanent deformation which eventually leads to an incremental collapse of the pavement. Note that even in the case of very small permanent deformations appearing in each cycle, its systematic accumulation could lead to an eventual failure of the structure due to excessive rutting. Whether a given material will experience progressive accumulation of permanent deformation, or whether this process will stop is therefore crucial for the performance prediction.  There are experimental evidences pointing out that permanent deformations quickly shake down, becoming the response basically resilient after a certain adaptation period \cite{allen74,lekarp98}.  A micro-mechanical observation of the permanent and resilient deformation will help to get insight into the mechanisms involved in the shakedown.

\begin{figure} [h]
  \centerline{
    \psfig{file=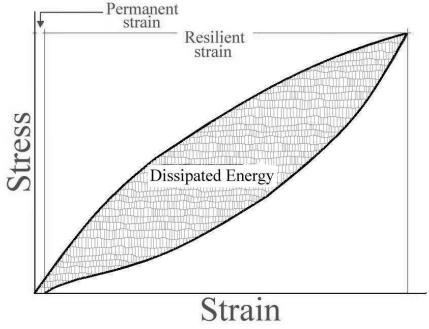,height=4.5cm,angle=0}}
    \caption{Visco-elastic stress-strain cycle. The resilient and permanent deformation is clearly identified.  At the end of the cycle, the system did not return to its original position, but accumulated a certain permanent deformation $\epsilon$. The energy dissipated in each cycle is the area enclosed by the loading and unloading curves.}
    \label{fig:resilient}
\end{figure}

Most of the theoretical approaches to UGM deformation try to identify the internal variables of postulated constitutive equations of the material, based on macro-mechanical observations of the response of soil samples in triaxial or biaxial tests. Shakedown theory, however, is concerned with the evolution of the plastic deformation in the material. It predicts that a structure is liable to show progressive accumulation of plastic strains under repeated loading if the magnitude of the applied loads exceeds a certain limiting value, the so-called {\it shakedown limit} or {\it limit load}. The structure is then said to exhibit an incremental collapse. On the other hand, if the loads remain below this limit, the growth of plastic deformations will eventually level off and the structure is said to have attained a state of shakedown by means of adaptation to the applied loads. Under these premises, four categories of material response under repeated loading can be distinguished \cite{collins00}:
\begin{itemize}
\item {\it Elastic} range for low enough loading levels, in which no permanent strain accumulation occurs.

\item {\it Elastic shakedown}. The applied stress is slightly below the plastic shakedown limit. The material response is plastic for a finite number of cycles, although the ultimate response is elastic.

\item {\it Plastic shakedown}. The applied stress is low enough to avoid a rapid incremental collapse. The material achieves a long-term steady state response with no accumulation of plastic strain, but hysteresis in the stress-strain plot.

\item {\it Incremental collapse}. The repeated stress is relatively large, so that plastic strain accumulates rapidly with failure occurring in the relatively short term.
\end{itemize}

The aim of this paper is to prove the existence of a shakedown regime in a simple model of UGM. This is done characterizing the material response to different loadings by means of the calculation of the dissipated energy throughout the experiment. A further description of the shakedown in the framework of the general elasto-plastic theory will be presented in the next section. The model used in our simulations is carefully described in Section $\ref{model}$, where the inter-particle interactions are also detailed. The results of the simulation of the system submitted to different loadings ranges are shown in Sections $\ref{shakedown-res}$  and Section  $\ref{plastic}$ . Finally, in Section $\ref{conclusions}$ the conclusions of this work are summarized. 

\section{Shakedown theory.}
\label{shakedown}
Shakedown analysis is basically an extension of limit analysis to the case of periodic loading \cite{koenig87}. It was introduced in the context of structural design to take into account that, in the case of repeated loads, an accumulation of small plastic deformations may occur in every cycle, that eventually might lead to a collapse of the structure. It was necessary, therefore, to develop methods able to estimate the {\it limit load}, beyond which failure occurs by either incremental collapse or fatigue. Below this load the structure will {\it shake down}, in the sense that plastic deformation ceases at some point or, more precisely, that the energy dissipated is bounded in time. Melan's pioneer works \cite{melan36} were completed by Koiter \cite{koiter60}, who first formulated a general kinematical shakedown theorem. Melan's and Koiter's theorems offer an lower and upper bond for the limit load, respectively. Their ideas are applied in \cite{sharp84} to a simple pavement model subjected to repeated moving loads, transforming the shakedown theorems into a linear programming problem, which is then numerically solved. 

Shakedown theory should be understood within the general framework of the classical theory of  elastoplasticity, where the existence of an elastic regime is postulated \cite{drucker52}. Elastoplastic constitutive laws consist of linear strain-stress relations that result in a overall incrementally non-linear relation. The basic assumption of these theories is that the stress state at any point of the material can be separated into two clearly different elastic and plastic parts: 
\begin{equation}
\sigma_{ij}(x,y)=  \sigma^E_{ij}(x,y) + \sigma^P_{ij}(x,y),
\label{eq:straindecomp}
\end{equation}
where  $\sigma^P_{ij}$ represents the residual stress due to the accumulated permanent deformation (see Figure $\ref{fig:resilient}$) and  $\sigma^E_{ij}$ is the elastic stress, related to the resilient strain $\epsilon^R_{ij}$ through the generalized Young's moduli $E_{ijkl}$:
\begin{equation}
\epsilon^R=E^{-1}_{ijkl} \sigma^E_{kl}.
\label{eq:hooke}
\end{equation}
This relation is usually known as the generalized Hooke's law.
In the space of stresses, there is a certain volume around the origin in which the system experiences an elastic deformation. The surface enclosing this area is a very special surface, called the {\it yield surface}. It is calculated in terms of the yield function $f(\sigma_{ij},\epsilon^P_{ij})$:
\begin{equation}
f(\sigma_{ij}, \epsilon^P_{ij})=0,
\label{eq:yield}
\end{equation}
The yield function $f$ is characteristic of the material. If it does not depend on the accumulated deformation, $\epsilon^P_{ij}$, the material is ideally plastic, whereas in any other case, it is said to suffer {\it hardening}. As hardening occurs, the size of the yield surface consequently changes, but in the case of isotropic hardening, neither the shape nor the orientation of the yield surface is altered.

When the stress state reaches the yield criterion ($\ref{eq:yield}$) the material undergoes plastic deformations, being the increments of plastic strain given by the normality rule:
\begin{equation}
d \epsilon^P_{ij}=  \Delta \Psi \frac{\partial Q}{\partial \sigma_{ij}}.
\label{eq:plastestr1}
\end{equation}
In this expression, the existence of the plastic potential function $Q$, to which the incremental strain vectors are orthogonal, has been assumed. $\Delta \Psi$ determines the magnitude of the plastic deformation. In the simplest case, the plastic potential $Q$ and the yield function are the same, i.e. $Q \equiv f$ (associative flow rule of plasticity). It will be assumed in the following that, during the plastic deformations, the yield surface translates in the stress space. This can be expressed:
\begin{equation}
f(\sigma_{ij} - \alpha_{ij}, \epsilon^P_{ij}) = 0,
\label{eq:yield1}
\end{equation}
where $\alpha_{ij}$ represents the total translation of the yield surface. Note that ($\ref{eq:plastestr1}$) and ($\ref{eq:yield1}$) are special cases of general anisotropic hardening, involving change in size, shape and orientation of the yield surface. Let us now consider the experiment shown in Figure $\ref{fig:resilient}$, in which the system reacts plastically to the imposed load. The response of the system is then discretized in $r$ linear segments in which the system behaves as described in equations ($\ref{eq:plastestr1}$) and ($\ref{eq:yield1}$). The total strain can still be decomposed in a plastic and an elastic component, as in ($\ref{eq:straindecomp}$), and the plastic strain rate can be determined,
\begin{equation}
\dot{\epsilon}_{ij}= \sum_{k=1}^{r} \dot{\Psi}^k \frac{\partial f_k}{\partial \sigma_{ij}},
\end{equation}
in terms of the $r$ yield surfaces $f_k$ and the material rates $\dot{\Psi}^k$, which respectively fulfill the following conditions:
\begin{equation}
\dot{\Psi}^k \geq 0, \; \dot{\Psi}^k f_k =0 \; \; \mbox{for} \; k=1, ...,r.
\end{equation} 
The yield condition has been then separated in $r$ parts, each of them given by:
\begin{equation}
f_k(\sigma_{ij}- \alpha_{ij}^k) \leq 0.
\end{equation}

Physically, it has been assumed that there exists in the domain of interest a series of yield surfaces, each defining a region in the total stress space of the system. Note that, as a consequence of this construction, the system will behave elastically if it is subjected to a low enough stress. Note also that surface $f_0$ is contained in surface $f_1$, which is contained in $f_2$, and so on... . In the experiments, the surfaces may deform and translate in the stress space. As soon as surface $f_k$ touches $f_{k+1}$, they will move together until they touch the next surface.  The shift of $f_{k}$, $\alpha_{ij}^k$, will be proportional to the plastic displacement in the set of yield surfaces, and the same reasoning can be applied to its rate,
\begin{equation}
 \dot{\alpha}_{ij}^k= \sum_{l=1}^{r} c^{kl} \dot{\Psi}^l \frac{\partial f_l}{\partial \sigma_{ij}},
\end{equation}
being $c^{kl}$ the strain-hardening coefficients. For this model, under the proper conditions \cite{koenig87}, Melan's static shakedown theorem assures that the total energy dissipated in any allowable stress path is bounded \cite{collins00}. Since direct experimental measure of the dissipated energy in the cycle is hardly possible, computer simulation seems to be the best tool for investigating the shakedown. In the particular case of unbound granular materials, Molecular Dynamics is the more appropriate method, for it solves the dynamics of the system after well established collision rules, and also because a detailed study of the micro-mechanics of the system and its relation to the macroscopic behavior of the material is possible.

Werkmeister {\it et al.} \cite{werkmeister01} have presented some experimental results on the shakedown of unbound granular material subjected to triaxial tests. They conclude that shakedown theory can be applied to study the formation of instabilities and degradation of UGM, only if some aspects of it are modified. The existence of three possible responses of the system is consequently proposed, depending on the load. For low loads, the system will definitely shakedown (behaving almost elastically), after a post-compaction transient in which plastic deformation is accumulated. On the contrary, for very high loads the material deforms boundless at an almost constant rate. The case of intermediate loads is somehow ambiguously treated by Werkmeister. Experimentally, however, it is clearly found that the plastic strain rate decreases to a low, nearly constant level for intermediate loads. A small residual linear increment of plastic strain is accumulated after each cycle.

\section{Model.}
\label{model}
 A polydisperse system of inelastic disks has been used to model the UGM. The chosen type of grain is an appropriate idealization of natural sands like gravel, in which the aggregate particles are smooth and round. The boundary conditions of the biaxial test will be reproduced. The system is bounded by four mobile walls that induce a certain stress state in the system.  The grains interact with each other through a viscoelastic force that will be explained in detail next. The dynamics of the system is solved by means of a MD algorithm.

\subsection{Contact forces.}
In order to calculate the forces, it is assumed that all the disks have a characteristic thickness $L$. The force between two disks is then written as  $\vec{F}=\vec{f} L$ and the mass of the disks as $M = m L$.  In any contact between two real grains, there is deformation in the impact region. In the simulation presented here the disks are supposed to be rigid, but they can overlap so that the force is calculated by means of this virtual overlap.

The contact force can be divided in two components, $\vec{f}^c=\vec{f}^e+\vec{f}^v$, where $\vec{f}^e$  and  $\vec{f}^v$ are the elastic and viscous contribution. The elastic part of the contact force also splits in two components, $\vec{f^e}= f^e_n \hat{n}^c + f^e_t \hat{t}^c$. Where $\hat{n}^c=(\vec{r}_i-\vec{r}_j)/ |\vec{r}_i-\vec{r}_j|$ is the unitary vector pointing from the center of mass of particle $j$ to particle $j$ and $\hat{t}^c=u_z \times \hat{n}^c$ ($u_z$ is the unitary vector in the direction perpendicular to the plane of motion of the disks).

\subsubsection{Normal elastic forces.}
 Given two overlapping disks $i$ and $j$ whose diameters are $d_i$ and $d_j$, respectively, the normal elastic force acting on them is $f^e_n=-k_n \frac{ \cal{A}}{d_i+d_j}$. Where $k_n$ is the normal stiffness and  $\cal{A}$ is the overlapping area. The overlapping area of these two particles can be easily calculated in terms of their diameters and the distance that separates their centers, $y_0$;
\begin{eqnarray}
{\cal{A}} = -2 y_0 x_0 + d_j^2 \left [ \frac{x_0}{d_j} \sqrt{1- \left ( \frac{x_0}{d_j} \right )^2} + \arcsin{(x_0/d_j)} \right ]+ \nonumber \\
 d_i^2 \left [ \frac{x_0}{d_i} \sqrt{1- \left ( \frac{x_0}{d_i} \right )^2} + \arcsin{(x_0/d_i)} \right ]~~~,
\end{eqnarray}
where the intersection point of both circles, $x_0$, has been introduced,
\begin{equation}
x_0^2=d_i^2- \left [ \frac{(d_i^2-d_j^2)+y_0^2}{2y_0} \right ].
\end{equation}
Note that the use of this model for the calculation of the elastic force is similar to the use of the overlapping length \cite{luding98c}, except for a renormalization factor in the elastic constant of the spring.

\subsubsection{Friction forces.}
In order to model the quasi-static friction force, the elastic tangential force is calculated using an extension of the method proposed by  Cundall-Strack \cite{cundall89}:   An elastic force   $f^e_t= -k_t \Delta x_t $ proportional to the elastic displacement is included at each contact. $k_t$ is the tangential stiffness of the contacts. The elastic displacement $\Delta x_t $  is calculated as the time integral  of the tangential velocity of the contact during the time that the  elastic condition  $|f_t|<\mu f_n$  is satisfied. Here $\mu$ is the friction coefficient. The sliding condition is imposed  keeping this force  constant when $|f_t|=\mu f_n$. The straightforward calculation of this elastic displacement is given by the time integral starting at the beginning of the contact
\begin{equation}
\Delta x^e_t=\int_{0}^{t}v^c_t(t')\Theta(\mu f^e_n-|f^e_t|)dt',
\label{friction} 
\end{equation}
where $\Theta$ is the Heaviside step function and $\vec{v}^c_t$ denotes the 
tangential component of the relative velocity $\vec{v}^c$ at the contact.
$\vec{v}^c$ depends on the linear velocity $\vec{v}_i$ and angular velocity 
$\vec{\omega}_i$ of the particles in contact according to:
\begin{equation}
\vec{v}^c=\vec{v}_{i}-\vec{v}_{j}-\vec{\omega}_{i}\times \frac{d_1}{2}\hat{n}^c
+\vec{\omega}_{j}\times \frac{d_2}{2}\hat{n}^c.
\end{equation}

\subsubsection{Damping forces.}
A rapid relaxation during the preparation of the sample and a reduction of the acoustic waves produced  during the loading is obtained if damping forces are included. These forces are calculated as $\vec{f}^v = -\nu m \vec{v}^c$. Here $m=m_i m_j/(m_i+m_j)$ is the relative mass of the disks in contact and $\nu$ is the coefficient of viscosity.  These forces introduce time dependence effects during the cyclic loading. Nevertheless, these effects can be arbitrarily reduced by increasing  the time of loading, so that the quasi-static approximation can be assumed.

\subsection{Sample preparation.}
The disks are initially randomly distributed into a rectangular box which is big enough for them not to overlap. Their radii distribution is Gaussian with mean value $\lambda$  and standard deviation $0.23 \lambda$. The interaction of the disks with the  walls of the box is implemented applying a normal visco-elastic force $f^w_n=-k_n\delta-m\nu_n v^c_n$ at each disk lodged in any of the walls. Here $\delta$ is the penetration length of the disk into the wall.  A gravitational field  $\vec{g}=g\vec{r}$ is also temporarily switched on, where $\vec{r}$ is  the vector connecting the center of mass of the assembly with the center of the disk. The activation of this gravity field produces homogeneous, isotropic distribution of disks.  After a certain time $t_0$, (which is defined below) a modulation in gravity is imposed such as $g = g_0+1/2(g_f-g_0)(1+cos(100\pi t/t_0))$ until the time $2 t_0$ in order to reduce the porosity. For $g_f=100 g_0$ ($g_0$ will be defined next), samples are obtained with packing fraction $0.841 \pm0.001 $.

The external load is imposed applying a force $\sigma_x W$ and $\sigma_y H$ on the  horizontal and  vertical walls, respectively. $W$  and $H$ are correspondingly the width and the height  of the sample. When the velocity of the disks vanishes, the gravity is switched off. A fifth-order predictor-corrector algorithm is used to  solve the equations of motion. The time scales used both for the compression and the cyclic loading are such that we can assume a quasi-static regime. This allows us to consider a static friction coefficient only.

It can be shown by dimensional analysis that the strain response depends only on a minimum set of dimensionless parameters: 1) the ratio $t_0/t_s$, between the period of  cyclic loading $t_0$ and the characteristic period of  oscillation  $t_s=\sqrt{k_n/\rho\lambda^2}$ (where $k_n$ is the normal stiffness of  the contacts and $\rho$  the density of the grains). 2) The ratio $t_r/t_s$ between the relaxation time  $t_r=1/\nu$ and the oscillation time. 3) the ratio $k_t/k_n$ between the stiffnesses. 4) the adimensional stress state ${\bf \sigma}/k_n$ and 5)  the friction  coefficient $\mu$. In our simulation the following values have been kept constant: $t_0 = 1000 t_s$, $t_r = 100 t_s$,  $k_t = 0.33 k_n$, $k_n=2$ MPa, $\mu=0.1$, $g_0=6.25 \times 10^{-8} k_n$, the initial pressure  $P_0=6 \times 10^{-4} k_n$, and the MD time step $t_s=2.5 \times 10^ {-6} s$. The imposed load, $\Delta \sigma$, has been the variable parameter used to investigate the response of the system. 
\begin{figure}
 \centerline{   
    \psfig{file=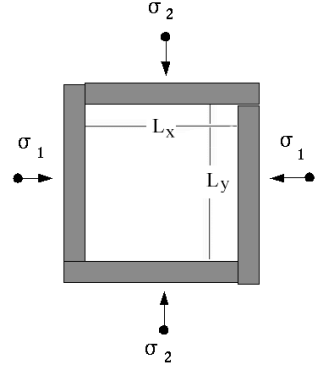,height=5.cm,angle=0}}
    \caption{Hambly's principle for biaxial test. The degrees of freedom of the walls allow to impose any pair of stresses $\sigma_1$, $\sigma_2$ to the system.}
    \label{fig:biaxial}
\end{figure}

\subsection{Biaxial test.}
The boundary conditions in the simulations are those of a biaxial test according to Hambly's principle (see Figure $\ref{fig:biaxial}$). In this experiment, the probe is compressed by four mobile walls which exert on the system a fixed stress $\sigma_1 \equiv \sigma_{xx}$ in the $x$ direction and $\sigma_2 \equiv \sigma_{yy}$ in the $y$ direction. The stress values can be changed and the corresponding changes in the principal components of the strain tensor $\epsilon_1$, $\epsilon_2$ can be measured. In our tests, the system is first homogeneously compressed with $\sigma_1=\sigma_2$. After the system has reached an equilibrium state under the pressure $P_0=\frac{\sigma_1+\sigma_2}{2}=\sigma_1$, the vertical stress is quasi-statically changed:
\begin{equation}
\sigma_2(t) = P_0 \left[ 1 + \frac{\Delta \sigma}{2} \left( 1- \cos \left( \frac{2 \pi t}{t_0} \right) \right) \right],
\end{equation}
where $t$ is the simulational time and $t_0$ is the period of the cyclic loading. It will be useful to define the deviatoric stress of this configuration, $S_{ij}$:
\begin{eqnarray}
\left(
\begin{array} {cc}
S_1(t) & 0\\
0 & S_2(t)
\end{array} \right)
 =
\left(
\begin{array}{cc}
\frac{\sigma_1(t)-\sigma_2(t)}{2} & 0\\
0 & \frac{ \sigma_2(t)-\sigma_1(t)}{2}
\end{array}
\right) ~~~~~~~~~ \nonumber \\
=
\left(
\begin{array}{cc}
- \frac{P_0 \Delta \sigma}{4} (  1 - \cos ( \frac{2 \pi t}{t_0} )  )  &0 \\
0 &\frac{P_0 \Delta \sigma}{4} (  1 - \cos ( \frac{2 \pi t}{t_0} ) )
\end{array} \right).
\end{eqnarray}

In the following experiments, the changes in the loading will be characterized by the parameter $\Delta \sigma$, which is directly related to the maximum value of the deviatoric strain. The response of the system will be characterized by the adimensional quantity $\gamma$, which is defined in terms of the deviatoric permanent strain. This, analogously to the deviatoric stress, is the difference of the strains in the principal directions. The permanent strains in the principal directions are:
\begin{eqnarray}
\epsilon_1(t)=\frac{L_x(t)}{L_x^0},\\
\epsilon_2(t)=\frac{L_y(t)}{L_y^0}.
\end{eqnarray}
Where $L_{x/y}(t)$ are the dimensions of the system at the time $t$, whereas $L_{x/y}^0$ are the dimensions at the beginning of the cyclic loading. Then, $\gamma$ is defined as: 
\begin{equation}
\gamma=\epsilon_2-\epsilon_1.
\label{eq:gamma}
\end{equation}
\begin{figure} [h]
\centerline{
    \psfig{file=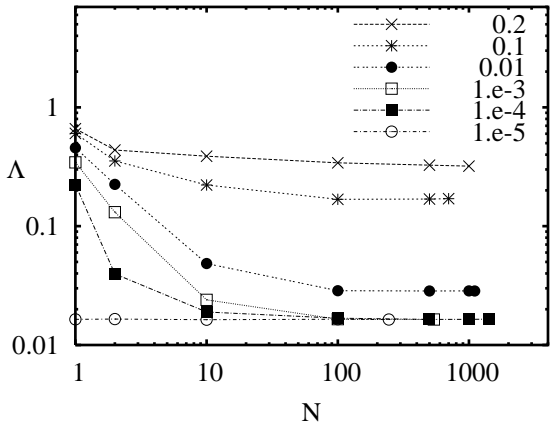,height=6.cm,angle=0}}
    \caption{Dissipated energy per cycle $\Lambda$ in the system for different loadings $\Delta \sigma$. After a transient in which energy is quickly dissipated, the system reaches a state in which $\Lambda$ is approximately constant. This value is higher for higher $\Delta \sigma$. There is however a critical value of $\Delta \sigma$ below which there is only viscous dissipation in the system.}
    \label{fig:elastic-disipa}
\end{figure}
\section{Description of the plastic shakedown.}
\label{shakedown-res}
According to shakedown theory, there are two possible fully elastic ranges in the deformation of the material. On one hand, a pure elastic response will be found for low enough loadings. On the other hand, elastic shakedown will come up if the load is slightly increased: The system will show some plastic deformation, that will eventually level off leading finally to an elastic response. 

The response of the system in the limit $\Delta \sigma \rightarrow 0$ will be discussed at this point. Within the framework of elastoplasticity, the material should then behave elastically, being the recoverable strain after a cycle a mathematical function of the stress. In the simplest case this relation should be linear, Eq.($\ref{eq:hooke}$). As pointed out in Figure $\ref{fig:resilient}$, the dissipated energy is the area of the cycle (C) in the stress-strain plane. Note that the total work applied to the system is the integral of the stress-strain curve during the loading (L). The ratio of these two magnitudes will help us to characterize the response of the system:
\begin{equation}
\Lambda=\frac{ \oint_{C} \sigma (\epsilon) d \epsilon}{\int_{L} \sigma (\epsilon) d\epsilon}.
\end{equation}

\begin{figure} [h]
\begin{tabular}{c}
  \psfig{file=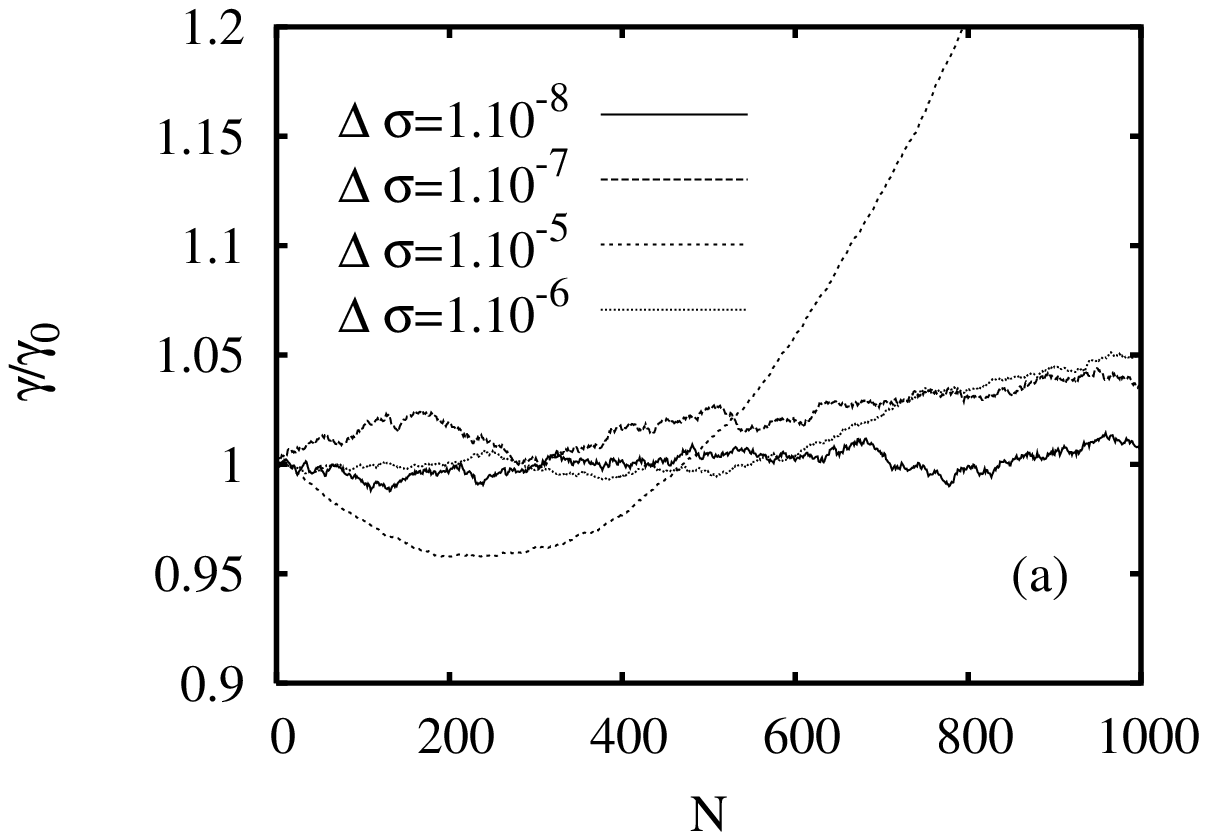,height=6.cm,width=8.cm,angle=0}\\
 \psfig{file=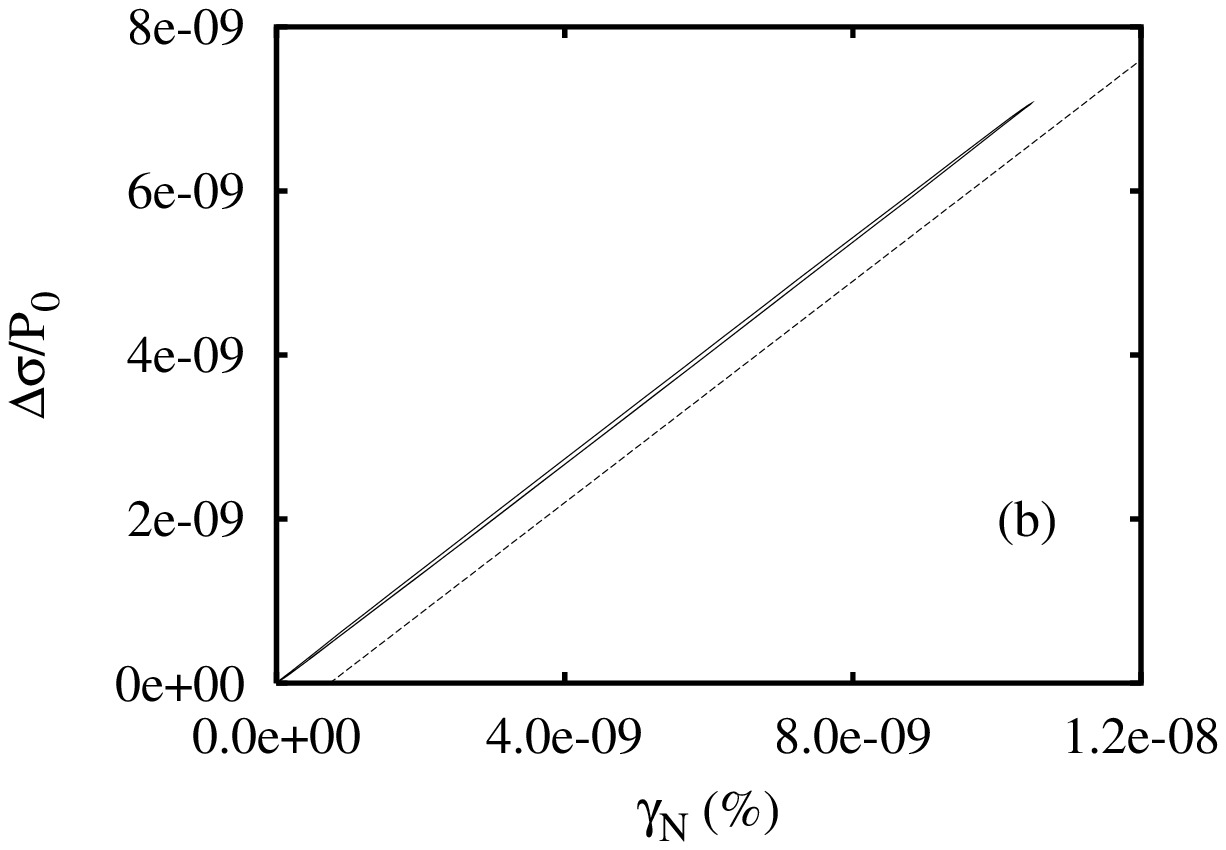,height=6.cm,width=8.cm,angle=0}          
\end{tabular}
    \caption{(a) Evolution of the permanent plastic strain in simulations with different deviatoric strains. For comparison, the values of the strain are scaled with the initial value $\gamma_0$. (b) Stress-strain plot for the case $\Delta \sigma = 10^{-8}$. The dotted line is plotted only as reference, showing a linear behavior. The details of the simulations are given in the text.}
    \label{fig:elastic}
\end{figure}

Figure $\ref{fig:elastic-disipa}$ shows the evolution of $\Lambda$ on time as $\Delta \sigma$ is changed. The system dissipates the most energy in the first cycles, reaching a state in which the energy dissipated in each cycle remains constant. This reflects the similarity of the stress-strain cycles and is associated to a certain periodicity of the sliding contacts \cite{alonso03}. The final value of $\Lambda$ clearly depends on $\Delta \sigma$. Note also that it takes longer for the system to reach this final stage if the loading is increased. In fact, if the loading is very low, the relaxation is apparently immediate. In that situation, the limiting value reached by $\Lambda$ is independent of the loading imposed. For $\Delta \sigma = 10^{-3}$ and  $\Delta \sigma = 10^{-4}$, the energy dissipated in the system converges to the curves corresponding to $\Delta \sigma <10^{-5}$. This decay is associated with the disappearance of sliding contacts in the sample. These results confirm the existence of a range of possible loadings for which the response of the UGM would be a plastic shakedown. The system has accumulated systematic plastic deformation at the beginning of the loading, but this process stops when the contacts cease sliding. In the limit of $\Delta \sigma \rightarrow 0$ one would expect that the dissipated energy per cycle also tends to zero. This is in fact what is found in the simulation, but Figure $\ref{fig:elastic-disipa}$ also shows that the dissipated energy tends to zero keeping the ratio $\Lambda$ constant. This is associated to the fact that no sliding contacts are found in the system for that range of excitations. The dissipation is therefore completely due to the viscosity. The energy dissipated through the viscosity in this model is directly related to the mean velocity of the particles. In the ideal quasi-static limit, this velocity would be zero. In our simulations, it can be reduced making $t_0$ bigger. For a finite $t_0$, however, no purely elastic regime can be found in this model.

Let us next investigate the deformation of the system in the viscous regime. Part (a) of Figure $\ref{fig:elastic}$ shows the evolution of the permanent strain in four different experiments all of them corresponding to very low load. In the lowest load case, $\Delta \sigma = 10^{-8}$, the permanent strain does not seem to have a defined trend, but on the contrary, it apparently wanders around the initial value, reflecting the existence of dissipation (Part (b) of Figure $\ref{fig:elastic}$). The deviations from the original value grow as $\Delta \sigma$ increases. For the highest load shown, a steady expansion of the system follows a small initial compression. In the intermediate cases, the strain remains approximately constant until $N=500$ in the case $\Delta \sigma = 10^{-7}$, whereas for $\Delta \sigma = 10^{-6}$, the change appears even earlier. In part (b) of the figure, the stress-strain cycle is plotted for the lowest loading and compared with a linear function. The existence of dissipation can be observed, although the ratio of dissipated energy and the total elastic work is $\Lambda=0.017$.

\section{Investigation of the plastic response.}
\label{plastic}
\begin{figure} [h]
  \begin{tabular}{c}
    \psfig{file=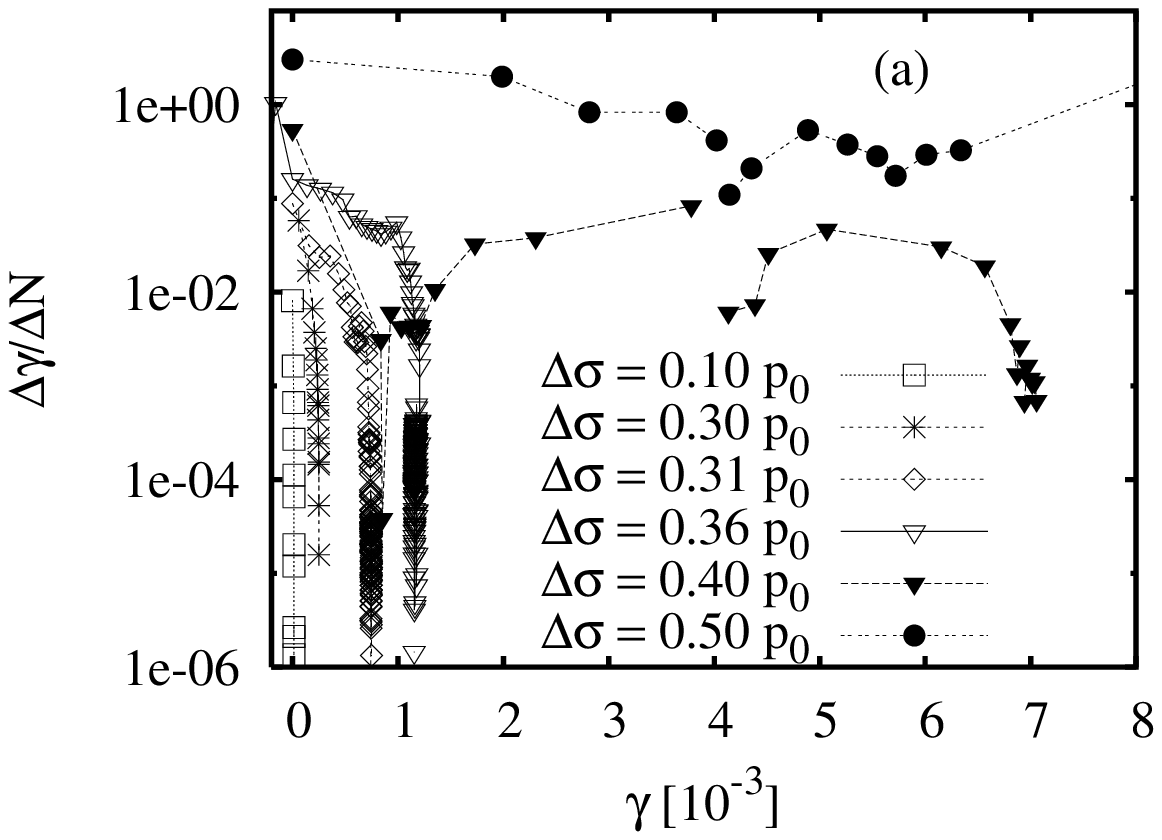,height=6.cm,width=8.cm,angle=0}\\
    \psfig{file=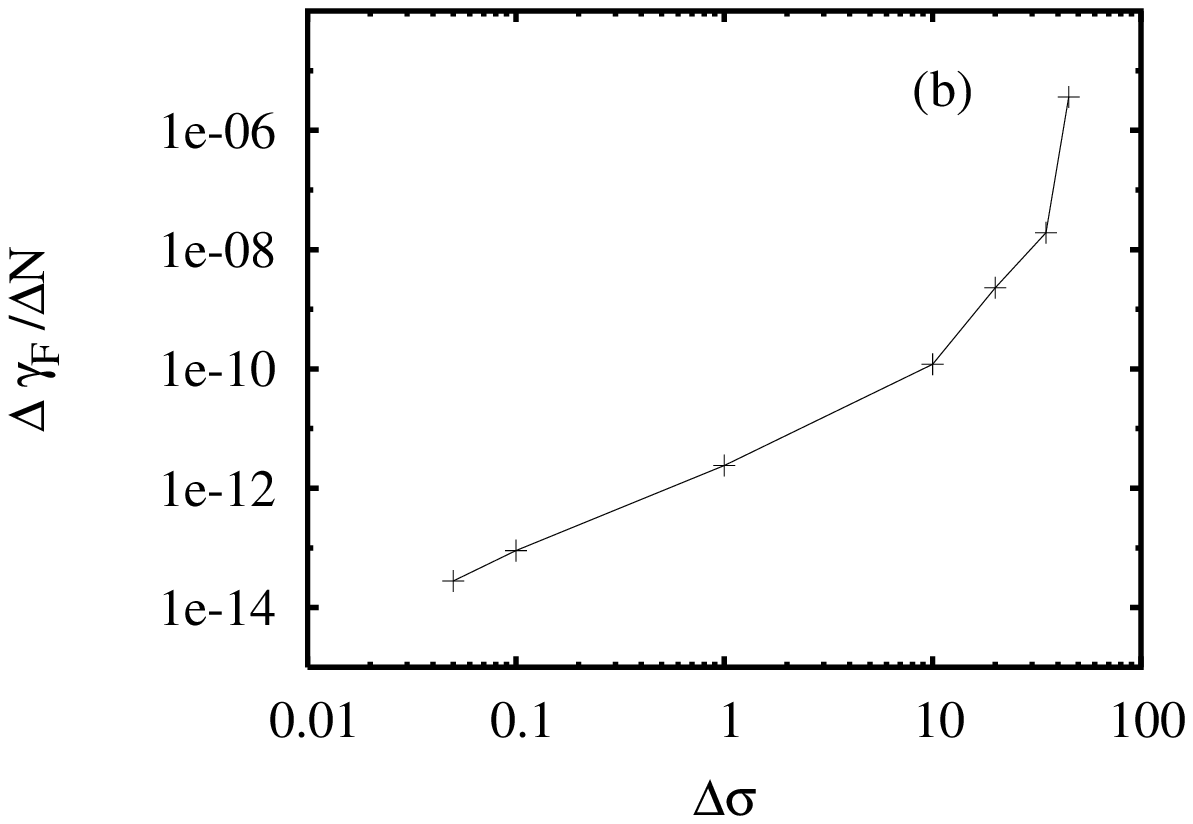,height=6.cm,width=8.cm,angle=0}
\end{tabular}
    \caption{(a) Strain rate versus the accumulated permanent strain for different $\Delta \sigma$. (b) Dependence of the final value of the strain rate ($\Delta \gamma_F / \Delta N$) on the loading $\Delta \sigma$. All the simulations correspond to the same initial configuration of $400$ disks.}
    \label{fig:strainrate}
\end{figure}

The plastic deformation induced in the granular material due to the cyclic loading has also been investigated. Given a system configuration and a certain confining pressure $P_0$ the response of the system to different deviatoric stresses is shown plotting the strain rate $\Delta \gamma/\Delta N$ versus the permanent strain $\gamma$, Figure $\ref{fig:strainrate}$. Part (a) of the figure helps to identify easily the existence of incremental collapse in the sample. For high loads, the strain rate does not diminish after each cycle, but remains approximately constant or even grows, reflecting the continuous accumulation of plastic deformation that the system is bearing. For low loading, however, the strain rate quickly decays to a certain almost constant value, implying a linear increase of the strain with the number of cycles. In both cases, a final constant value of the strain rate can be measured, $\Delta \gamma_F / \Delta N$. Note that the existence of a shakedown of the excitations cannot be consequently inferred from this graph. On the contrary, this ratcheting regime is associated with a periodic behavior of the sliding contacts and has been already reported both in a MD simulation of a polygonal packing \cite{alonso03} and experimentally \cite{werkmeister01}. The final permanent deformation is higher for higher applied loads. This is related to the dependence of the strain rate on $\Delta \sigma$ (Figure $\ref{fig:strainrate}$ (b)). The strain rate is higher as the loading is increased. The incremental collapse is clearly associated to an unbound growth of the strain rate.

These results are compatible with experimental measurements on a triaxial test by Werkmeister et al. \cite{werkmeister01}, even though no grain crushing is included in our model. This is probably the reason why the incremental collapse seems to be weaker in the simulation as it is in the experiment. 

\begin{figure} [h]
  \centerline{
    \psfig{file=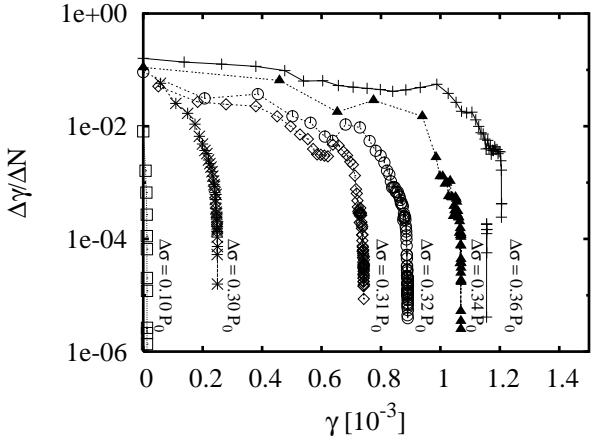,height=6.cm,angle=0}}
    \caption{Decrease of the strain rate through the cyclic loading to a constant value in the ratcheting regime.}
    \label{fig:strainrate2}
\end{figure}

The evolution of the strain rate for intermediate loadings is worth further discussion. Figure $\ref{fig:strainrate2}$ shows the response of the system for different excitations below the incremental collapse. Let us remark first the existence of two different regimes in the response of the material. At the beginning of the loading process, the system reacts accumulating deformation at a (relatively) high rate. This regime will be called {\it post-compaction}. After the post-compaction, the material has adapted to the new situation and the strain rate is very low and almost constant. This final stage is the ratcheting regime, in which the material accumulates deformation linearly. Within the loading cycles the system behaves quasi-periodically. The strain growth is linked to a small compression of the system. As $\Delta \sigma$ is increased, the {\it duration} of the post-compaction (in terms of the number of cycles) is longer. This is related to the existence of transient states, in which the strain accumulates at a constant rate for some cycles, before reaching its final level. This intermediate state is clearly observed in cases $\Delta \sigma =0.31,0.32$ and $0.34$. The strain rate decays to a fixed value for some time, but then the material reacts and an abrupt change on the strain rate leads to the final relaxation. This phenomenon is less clear as $\Delta \sigma$ increases and the strain is accumulated at an almost constant rate during the post-compaction. In this situation, if $\Delta \sigma$ is big enough, the system can even break, as it is noticeable in the case $\Delta \sigma =0.36$ in the figure. During the experiment, a sudden compression of the system occurs just after the post compaction, and previous to the complete adaptation of the material.

\begin{figure} [h]
\begin{tabular}{c}
\psfig{file=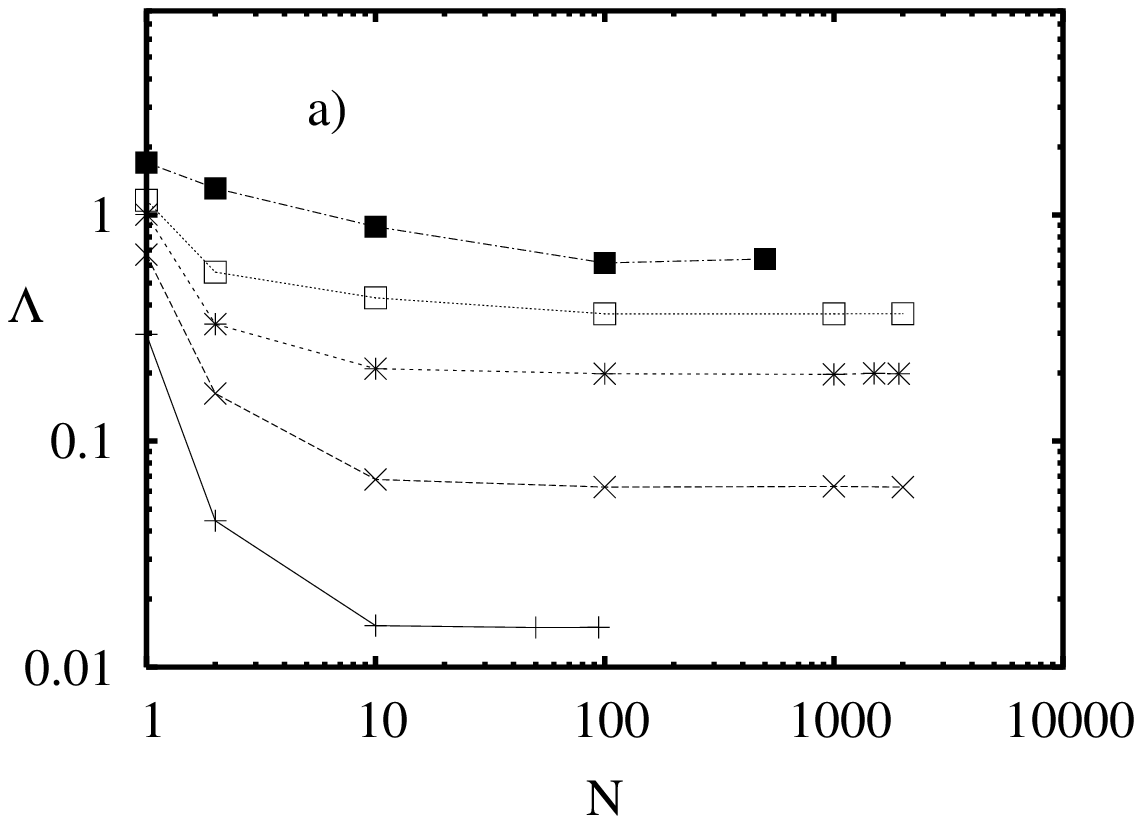,height=6.cm,width=8.cm,angle=0} \\
\psfig{file=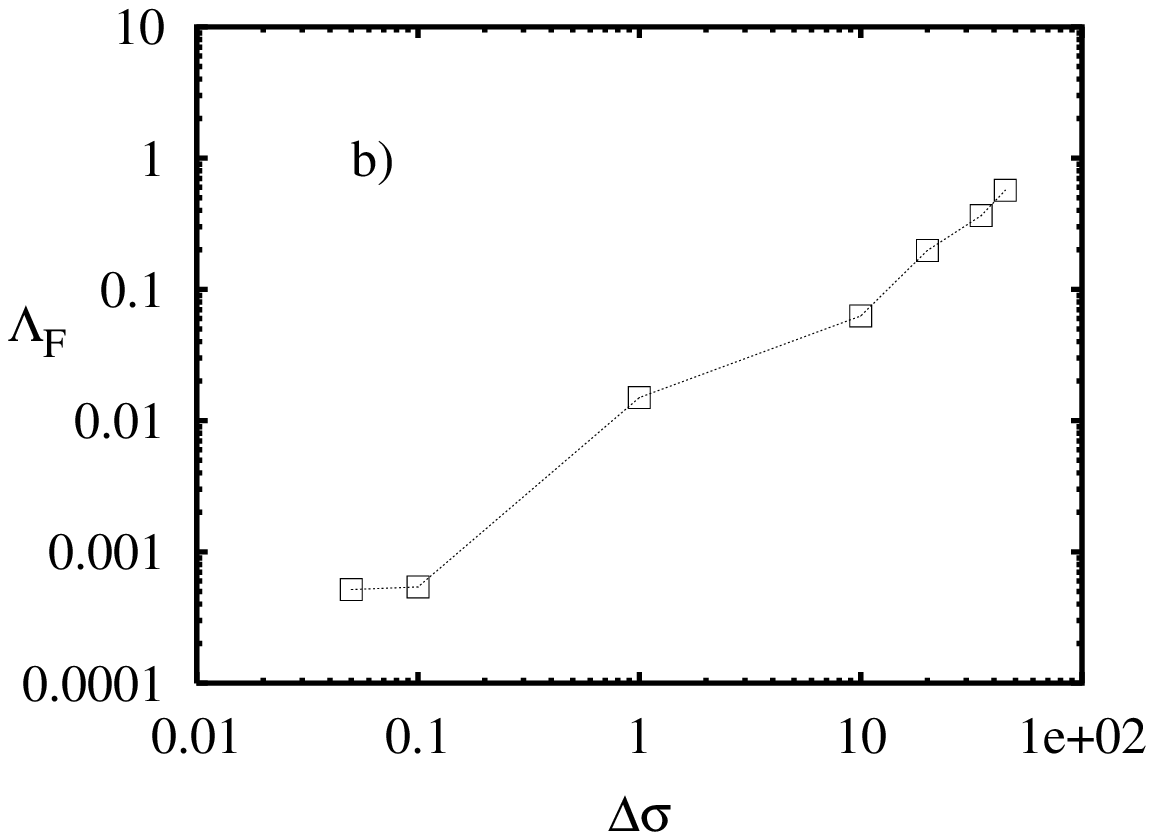,height=6.cm,width=8.cm,angle=0}          
\end{tabular}
    \caption{(a) Evolution of the scaled dissipated energy $\Lambda$ in time for different values of $\Delta \sigma$. The plotted values correspond to (from bottom to top) $\Delta \sigma=0.01$, $\Delta \sigma=0.10$, $\Delta \sigma=0.20$,  $\Delta \sigma=0.35$, and  $\Delta \sigma=0.45$. (b) Final dissipated energy ratios. Simulation details are the same as in Figure $\ref{fig:strainrate}$.}
    \label{fig:energy}
\end{figure}

\begin{figure} [h]
  \centerline{
    \psfig{file=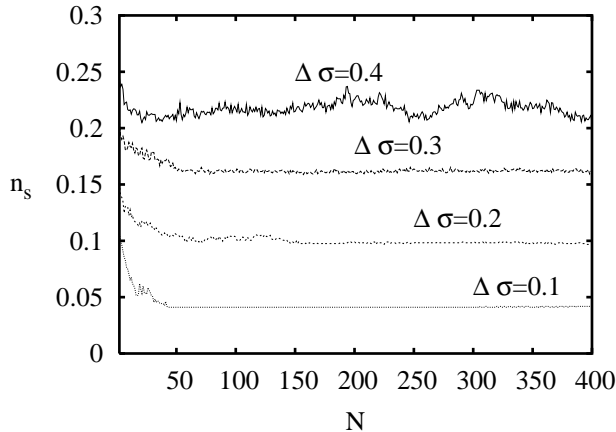,height=6.cm,angle=0}}
    \caption{Decay of the maximum relative number of sliding contacts $n_s$ in the post-compaction phase for the different loadings of Figure $\ref{fig:strainrate}$.}
    \label{fig:nslid}
\end{figure}

Extending the results of Figure $\ref{fig:elastic-disipa}$, a similar study on the dissipated energy in shown in Figure $\ref{fig:energy}$ for higher loadings. According to the shakedown theorems, none of the material to which these curves correspond would shake down, for the total dissipated energy is not bounded. The dissipated energy increases as $\Delta \sigma$ also increases. For very high loads, in fact, the dissipated energy can be even higher than the work exerted on the material ($\Delta \sigma =0.45$ of Figure $\ref{fig:energy}$). This implies an overall modification of the structure, including of course the destruction and creation of contacts. Since all the curves have a plateau for a large enough number of cycles, a measure of the final value, $\Lambda_F$, can be worked out. This is shown in part (b) of Figure $\ref{fig:energy}$. Above the shakedown limit, the increase of dissipated energy with $\Delta \sigma$ is potential in the range of parameters covered by our simulations. For very low $\Delta \sigma$ the system approaches the plastic shakedown regime, in which $\Lambda_F$ is independent of $\Delta \sigma$.
\begin{figure} [h]
  \centerline{
    \psfig{file=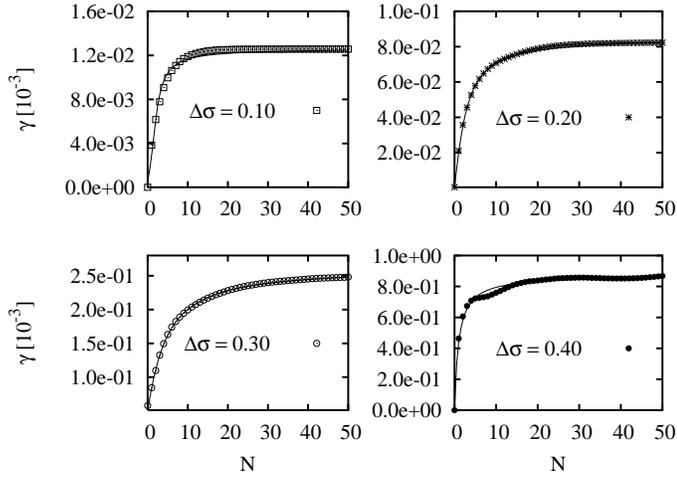,height=6.5cm,angle=0}}
    \caption{Fit of the strain-rate vs. strain curves in the post-compaction phase to a simple law (see text), to some of the curves plotted in Figure $\ref{fig:strainrate2}$.}
    \label{fig:fit}
\end{figure}
If the applied load is such that no incremental collapse is observed in the response, the post-compaction regime is characterized by a growth of the accumulated strain with $N$ as $P(N) \propto arctan{(b N+c)}$, being $a$, $b$, and $c$ adjusting parameters. Figure  $\ref{fig:fit}$ shows the fit of these data to the corresponding best-fit curves. This dependence is consequence of a simple second order polynomial relationship between the strain rate and the accumulation of strain. Only in case $\Delta \sigma =0.40$, some deviation is observed at the beginning of the loading, showing that this simple approximation is not valid for high loads. The behavior of the plastic strain in the post-compaction is related to a decay of the relative number of sliding contacts in the early phases of the loading. Due to the cyclic loading, a periodic sliding of the contacts is forced in the sample. The maximum number of sliding contacts in one cycle, $N_s$, decays to a nearly constant value in the first stages of the post-compaction. Figure $\ref{fig:nslid}$ shows the behavior of $N_s$ scaled with the total number of contacts: $n_s$. The final value of $n_s$ is clearly larger for the highest loadings, and so is its variability. For $\Delta \sigma=0.1$ the number of sliding contacts remains practically unaltered during the ratcheting. For $\Delta=0.4$, however, this number changes continuously from one cycle to the next. This mainly reflects the changes in the number of sliding contacts in the sample, although also the creation and destruction of contacts between particles is responsible for the changes in $n_s$.

\begin{figure} [h]
  \centerline{
    \psfig{file=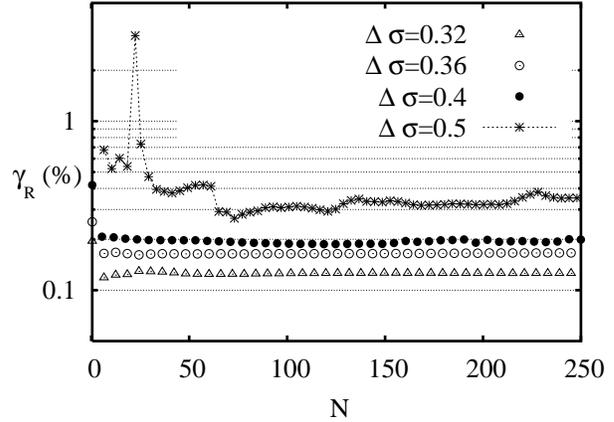,height=6.cm,angle=0}}
    \caption{Evolution of the resilient strain throughout the simulation for some of the systems included in Figure $\ref{fig:strainrate2}$.}
    \label{fig:resilientstrain}
\end{figure}

The evolution of the resilient strain will also help to get an insight into the mechanisms involved in the deformation. It is known in soils mechanics that the resilient strain changes only slightly through the simulation \cite{allen74}. It can be inferred from Figure $\ref{fig:resilientstrain}$ that the adaptation time of the system depends on the imposed loading. For large loadings, the resilient strain is still changing from one cycle to the next, after $200$ cycles. For lower excitations, however, the resilient strain does not change after 20 or 30 cycles. In the ratcheting regime, the stress-strain cycle maintains its shape and the resilient modulus remains therefore constant. In the incremental collapse, however, the resilient strain varies, reflecting the changes in the configuration of the sample. The changes, being minor for moderate deviatoric stresses ($\Delta \sigma=0.4$), become obvious when $\Delta \sigma$ is increased.

\section{Conclusions.}
\label{conclusions}
 A given material at rest or equilibrium under fixed stress conditions shows a certain configuration of contacts between particles and forces. When a cyclic loading is imposed, the system reacts by changing its configuration. In a continuous and gradual increase of the loading amplitude $\Delta \sigma$, the material will start by trying to change the force skeleton without altering its configuration of contacts \cite{mcnamara04}. Negligible creation or destruction of contacts will occur and, for low enough excitations, no sliding contacts can be found in the simulation. After one cycle, however, the particles will recover its original state only approximately because some energy will have been dissipated, i.e. due to viscosity. At this range of loadings, the dissipated energy is independent of the loading and, basically, does not change from one cycle to another. No pure elastic response is therefore found in the system, and the possibility of elastic shakedown is consequently also discarded. It is possible, however, that the material reaches a state in which there is no further systematic accumulation of permanent stress (plastic shakedown). 

For higher loadings, the energy input is first quickly dissipated by a re-arrangement of the sliding contacts of the system, the so-called post-compaction. The dissipated energy per cycle relaxes then to a stationary value dependent mainly on the loading, but also on characteristics of the grains such as the friction or the stiffness of the contacts. In this stage, the system evolves quasi-periodically. The sliding contacts show a periodic behavior within the cycles, although there is a linear increase of the deviatoric strain and an overall compression of the system. This reflects the fact that the system is still not dissipating all the energy it should. It is therefore expected that the material evolves on a much longer scale to a final shakedown state in which all the energy supplied to the system is dissipated. This process may take a longer time in the simulation than in the real experiment, where more dissipative mechanisms exist.

If the loads imposed are high enough, there is no possibility for the system to re-arrange to the new situation and the post-compaction is substituted by an incremental collapse. The energy that the system is not able to dissipate in its configuration, is used to modify its shape.

The stress-strain relations have been carefully studied in this paper, both as the mean to calculate the dissipated energy in each cycle, but also as the best way of characterizing the state of the system. In this respect, it has been proven that, for non-collapsing materials, the basic assumption expressed in ($\ref{eq:straindecomp}$), allows to fully characterize any stress state of the system in term of the strain rate, the resilient modulus and Poisson's ratio. Although there exist a large number of valuable models predicting the dependence of these resilient parameters on the stress \cite{lekarpb00}, little is known about how they are affected by material constants such as friction or the stiffness of the contacts. This dependencies, however, are required in order to predict the behavior of the UGM within a certain structure, and will be presented elsewhere.

\section{ACKNOWLEDGMENTS}

The authors want to aknowledge G. Gudehus's clarifying ideas on the granular ratcheting. Special thanks to Fernando Alonso-Marroq\'{\i}n for his help, support and valuable discussions. This work was possible thanks to the support of EU project {\it Degradation and Instabilities of Geomaterials with Application to Hazard Mitigation} (DIGA) in the framework of the Human Potential Program, Research Training Networks (HPRN-CT-2002-00220).

\bibliography{../granulates.bib}
\end{document}